\documentclass{iau}
\usepackage{graphicx}
\usepackage{natbib}

\title[IAUS291.~~The spin evolution of young pulsars] 
{The spin evolution of young pulsars} 

\author[C.~M.~Espinoza]  
{Crist\'obal M.~Espinoza$^1$}

\affiliation{$^1$Jodrell Bank Centre for Astrophysics, \\School of Physics and Astronomy, The University of Manchester, M13 9PL, UK.
\\ email: {\tt cme@jb.man.ac.uk} }

\pubyear{2012}
\volume{291}  
\jname{\mbox{Neutron Stars and Pulsars: Challenges and Opportunities after 80 years}}
\editors{J.~van Leeuwen, ed.}

\begin{document}
\maketitle

\begin{abstract}
 The current understanding of the spin evolution of young pulsars is reviewed through a compilation of braking index measurements.
 An immediate conclusion is that the spin evolution of all pulsars with a measured braking index is not purely caused by a constant magnetic dipole.
 The case of PSR J1734-3333 and its upward movement towards the magnetars is used as a guide to try to understand why pulsars evolve with $n<3$.
 Evolution between different pulsar families, driven by the emergence of a hidden internal magnetic field, appears as one possible picture.
  
\end{abstract}


The spin frequency ($\nu$) of all pulsars is observed to decrease and the rate of this spin-down ($\dot{\nu}$) is commonly related to the magnitude $B$ of a dipolar magnetic field rotating with the star.
The spin-down of pulsars can be characterised by the braking index $n$, which is defined via $\dot{\nu}\propto\nu^n$ and, provided we are able to detect the long-term second time-derivative of the spin frequency $\ddot{\nu}$, it can be measured by
\begin{equation}
n=\frac{\nu\ddot{\nu}}{\dot{\nu}^2} \quad.
\end{equation}

Another way to visualise the spin evolution of pulsars is by using a period--period-time-derivative plot with all known pulsars (the $P$--$\dot{P}$ diagram, Fig. \ref{fig:ppdot}).
As their periods increase, pulsars move to the right in this diagram, with a slope $2-n$ determined by the dominant braking mechanism in operation.
In particular, for braking due to a constant magnetic dipole we expect $n=3$ and pulsars to move to the right and downward with a slope of $-1$.  
To date, the few measurements available indicate $n<3$ (i.e. {\sl slope} $<-1$).


Commonly, the age of a pulsar is assumed to be equal to the so-called characteristic age, $\tau_c=\nu/2\dot{\nu}$, under the assumption that the pulsar evolved with a constant $n=3$ and that the initial spin frequency was significantly larger than the current value.
We know that young pulsars evolve with $n<3$, hence $\tau_c$ is a poor age estimator. 
More braking index measurements and a better understanding of the spin-down physics are essential to track the rotational history of pulsars and estimate their real ages.

The $P$--$\dot{P}$ diagram is populated by various families of pulsars which exhibit different emission or rotational properties and/or are found in different astrophysical contexts, like  binary systems, supernova remnants, etc. \citep[cf.][]{kas10}.
A better understanding of the movement of pulsars on the diagram will offer insights on the possible relationship between these families \citep{kas10,elk+11}, which is essential to understand the Galactic population of neutrons stars \citep{kk08}.

\section{PSR J1734$-$3333}
PSR J1734$-$3333 is a radio pulsar rotating with a period $P=1.169$\,s and with a rather high period derivative $\dot{P}=2.3\times10^{-12}$ implying $B=5\times10^{13}$\,G \citep{mhl+02}, a value comparable to the inferred magnetic field of some magnetars (Fig. \ref{fig:ppdot}).
There are no glitches detected for this pulsar and its levels of timing noise are relatively low.
By using 13.5 years of data from the Jodrell Bank Observatory and the Parkes telescope, \citet{elk+11} measured $\ddot{\nu}$ for this pulsar and calculated $n=0.9\pm0.2$. 
Therefore, the spin-down of PSR J1734$-$3333 is not purely caused by a constant dipole magnetic field.
Its movement in the $P$--$\dot{P}$ diagram is upwards and with a slope of at least $0.9$. 
If the physics responsible for this evolution remains the same, PSR J1734$-$3333 will be located among the magnetars, with a period of $8$\,s, in less than $30$\,kyr \citep{elk+11}.

\section{Other braking index measurements}
Braking index measurements have been possible only for a few pulsars. 
The main difficulty is found on the detection of the secular $\ddot{\nu}$ caused by the spin-down mechanism.
The effects of $\ddot{\nu}$ in the data, in addition to be very small, are  obscured by the effects of glitches and timing noise.
In this section we collect most $n$ measurements dividing the sample of pulsars in two main groups, according to their spin-down behaviour.

\begin{figure}[b]
 \begin{center}
  \includegraphics[width=2.6in]{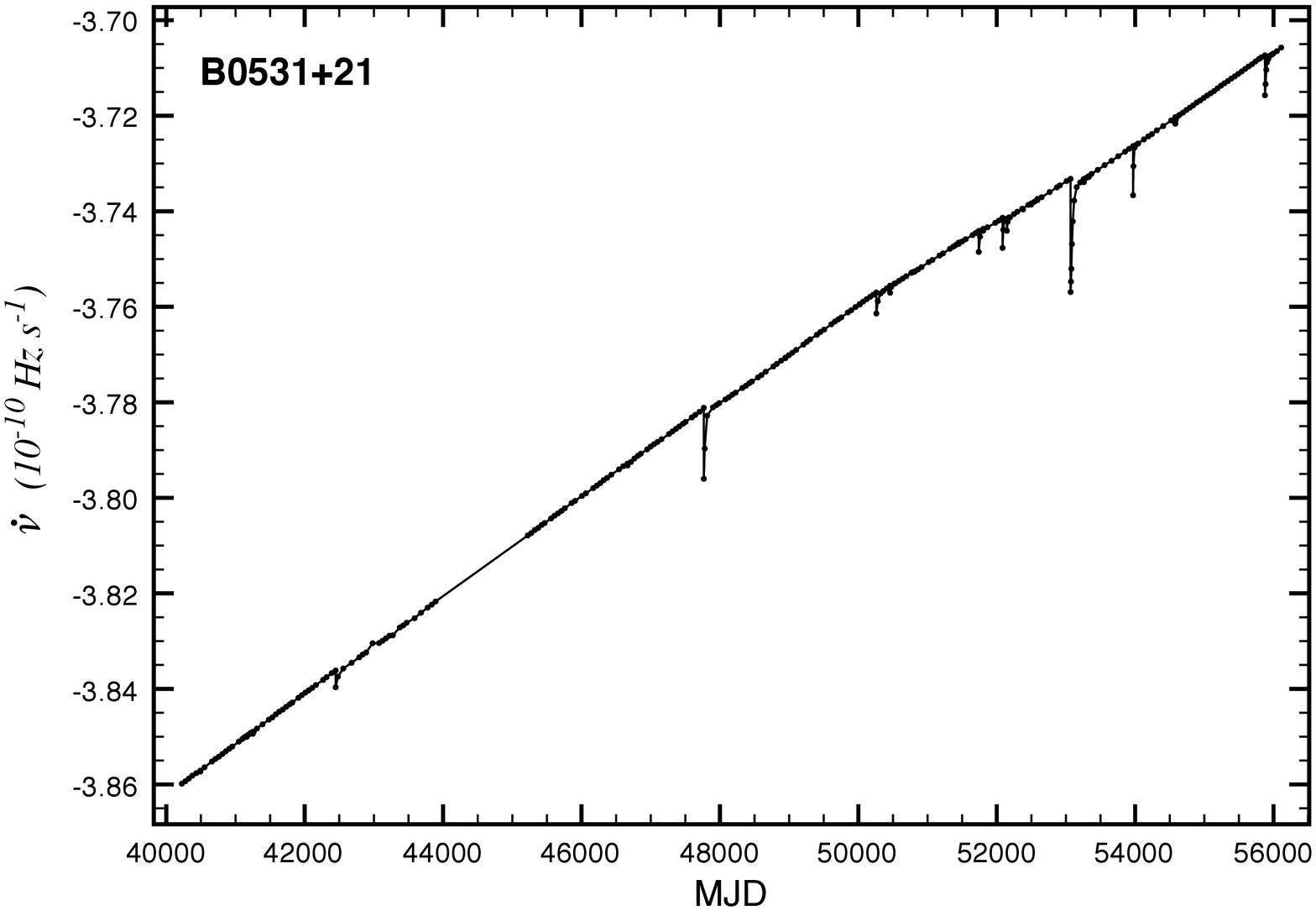}
  \includegraphics[width=2.6in]{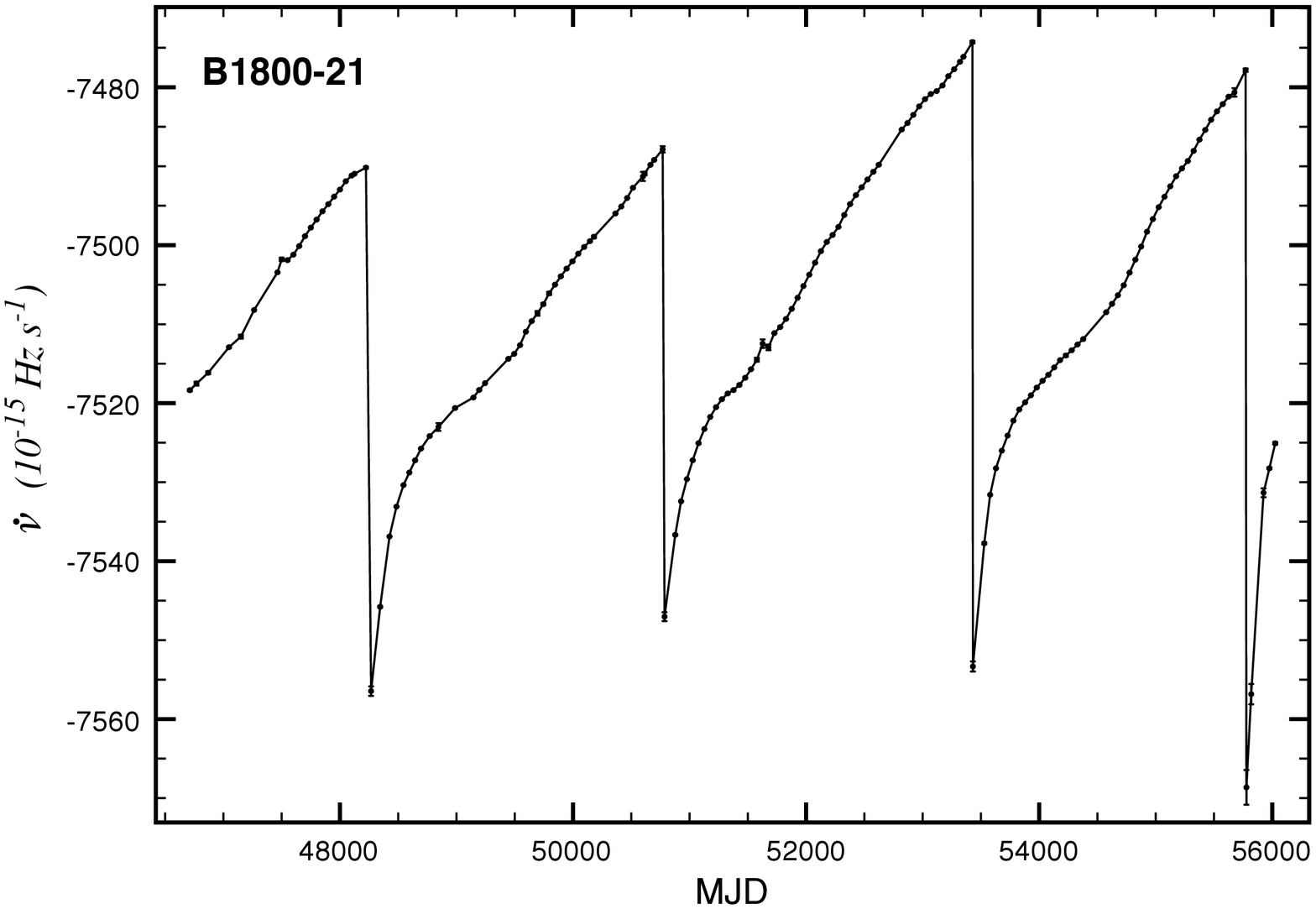}
  \caption{The $\dot{\nu}$ time-evolution of the Crab pulsar (PSR B0531+21) over more than $40$\,yr (left) and of PSR B1800$-$21 over about $25$\,yr (right). 
  Glitches appear as negative jumps.} 
  \label{fig:nudot}
 \end{center}
\end{figure}

\subsection{The case of very young pulsars}
Very young pulsars have $\tau_c$ values around $1$\,kyr and are associated to supernova remnants of similar ages.
They present rather linear $\dot{\nu}$ time-evolutions and the secular $\ddot{\nu}$ is easy to detect via coherent timing or by just measuring the slope of the $\dot{\nu}$ curve.
Glitches in these pulsars appear not to have a dramatic effect on the long-term linear behaviour of $\dot{\nu}$ 
(see, for example, the left panel in Fig. \ref{fig:nudot} and the $\dot{\nu}$ evolution of PSR J1119$-$6127 in Fig. 10 of \citet{wje10}).
The braking index measurements for these pulsars have all been performed via coherent timing and can be found on the left part of Table \ref{table}.
We include in this group a recent measurement for PSR J1833$-$1034.

\subsection{The case of Vela-like pulsars}
The braking index of the Vela pulsar (PSR B0833$-$45) was measured by \citet{lpgc96} using a particular method to overcome the presence of large glitches.
Other pulsars with similar rotational properties to the Vela pulsar also exhibit  large glitches of comparable size, occurring at semi-regular time intervals. 
It is this regularity what allowed \citet{lpgc96} to track the long-term $\dot{\nu}$-evolution of the Vela pulsar and measure $\ddot{\nu}$. 
We have updated the method and studied the $\dot{\nu}$ evolution of the Vela pulsar (now including 14 glitches) and the three Vela-like pulsars PSRs B1757$-$24, B1800$-$21 and B1823$-$13 \citep{e+12}.
As preliminary results, here we report rather low braking indices for all these four pulsars (Table \ref{table}).
Table \ref{table} also includes a braking index for PSR J0537$-$6910, based on a rough $\ddot{\nu}$ estimate performed over $\dot{\nu}$ data including a number of large glitches and an upper limit for PSR J1744$-$2958.

\begin{figure}[b]
 \begin{center}
  \includegraphics[width=2.8in]{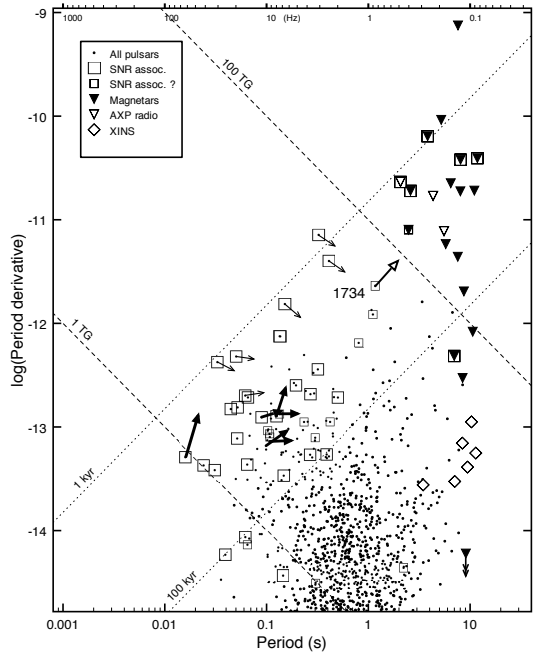} 
  \caption{The $P$--$\dot{P}$ diagram for pulsars with $\dot{P}>1.65\times10^{-15}$. 
  Arrows indicating the position after one $\tau_c$ are plotted for the pulsars in Table \ref{table}. 
  Thick, closed arrows are for Vela-like pulsars, the open arrow is for PSR J1734$-$3333 and the thin arrows are for the very young pulsars.
Constant $B$ lines are segmented and constant $\tau_c$ lines are dotted.} 
    \label{fig:ppdot}
 \end{center}
 \end{figure}

\begin{table}
  \begin{center}
  \caption{Measured braking indices.}
  \label{table}
    {\scriptsize
  \begin{tabular}{lllllcl}
  \hline
  \multicolumn{1}{l}{Pulsar} & \multicolumn{1}{c}{$n$} &  \multicolumn{1}{l}{Ref.} & &
    \multicolumn{1}{l}{Pulsar} & \multicolumn{1}{c}{$n$} &  \multicolumn{1}{l}{Ref.} \\
    \hline
B0531$+$21   & 2.51(1)  & \cite{lps93}  & & J0537$-$6910 & $-1.5$   & \cite{mmw+06} \\
B0540$-$69   & 2.140(9) & \cite{lkg+07} & & B0833$-$45   & $1.7$& \cite{e+12}   \\
J1119$-$6127 & 2.91(5)  & \cite{wje10}  & & J1734$-$3333 & 0.9(2)   & \cite{elk+11} \\
B1509$-$58   & 2.839(1) & \cite{lkg+07} & & J1747$-$2958 & $<1.3$   & \cite{hgc+09} \\
J1833$-$1034 & 1.857    & \cite{rgl12}  & & B1757$-$24   & $-1$     & \cite{e+12}   \\
J1846$-$0258 & 2.65(1)  & \cite{lkg+07} & & B1800$-$21   & $2$      & \cite{e+12}   \\
             &          &               & & B1823$-$13   & $2$      & \cite{e+12}   \\
\hline
  \end{tabular} }
  \end{center}
  \end{table}

\section{Discussion and questions for the future}
All 13 pulsars in Table \ref{table} exhibit $n<3$. 
This result implies that the spin-down of young pulsars is not driven by an electromagnetic torque caused by a constant magnetic dipole.
Either the dipole is changing with time, the braking is produced by higher order multipoles or there are other processes that efficiently compete with the electromagnetic torque, like stellar winds, other magnetospheric processes or internal processes related to superfluid dynamics \citep{br88,mic69,lst12,ha12}.

The low $n$ of PSR J1734$-$3333 has been attributed to an increase of the dipole component of its magnetic field \citep{elk+11}.
Simulations show that the dipole's magnitude could increase as a result of the re-emergence of a stronger magnetic field buried by hypercritical accretion occurred immediately after the supernova explosion \citep{mp96,vp12}.
If this is a common situation among young pulsars, different strengths and submergence conditions of the internal magnetic field, together with different initial periods, could produce the low braking indices measured on young pulsars, the properties of CCOs \citep{ho11,vp12} and also offer a new genesis for the magnetars \citep{lyn04,elk+11}. 
In this picture, CCOs, young radio pulsars and magnetars are all one single family, conveniently reducing the total number of neutron stars in the Galaxy \citep[c.f.][]{kk08}.

We acknowledge, however, that the actual situation might be more complicated because
the spin-down of pulsars is probably the result of a superposition of various processes.
The values in Table \ref{table} are represented in the $P$--$\dot{P}$ diagram in Fig. \ref{fig:ppdot} by arrows, which indicate the position pulsars will have after one $\tau_c$\footnote{Excepting PSR J0537$-$6910, for which only 2,000 yr were used.}.
In general, very young pulsars have negative slopes and Vela-like pulsars and PSR J1734-3333 have positive slopes.
Are the pulsars in these two groups irreconcilably different?
Or, will very young pulsars evolve into Vela-like pulsars?
What sort of processes could produce this transformation?

It is practically impossible to discriminate from timing data which is the exact spin-down mechanism operating on a given pulsar.
Hence, it is unclear whether the spin evolution of very young pulsars is caused by the same process as the other pulsars or not.
In the future, it will be important to study the efficiency of the different braking mechanisms and understand how they compete with each other; what kind of observable signatures we might expect from each of them and how they evolve as pulsars age.

\end{document}